\documentclass[journal]{IEEEtran}
\IEEEoverridecommandlockouts
\usepackage{subfigure}
\usepackage{subcaption}
\usepackage{stfloats}
\usepackage{cite}
\usepackage{amsmath,amssymb,amsfonts}
\usepackage{algorithmic}
\usepackage{graphicx}
\usepackage{textcomp}
\usepackage{xcolor}
\usepackage{multirow}
\usepackage{orcidlink}   % REMOVE THIS LINE
\usepackage{hyperref}
\def\BibTeX{{\rm B\kern-.05em{\sc i\kern-.025em b}\kern-.08em
    T\kern-.1667em\lower.7ex\hbox{E}\kern-.125emX}}
\begin{document}

\title{\fontsize{18}{20}\selectfont \textbf{XR-NPE}: High-Throughput Mixed-precision SIMD \\ \textbf{N}eural \textbf{P}rocessing \textbf{E}ngine for E\textbf{x}tended \textbf{R}eality Perception Workloads\\

\thanks{
\textsuperscript{\textdagger}Both authors contributed equally to this work.
This work was supported by the Special Manpower Development Program for Chip-to-Start-Up (SMDP-C2S), the Ministry of Electronics and Information Technology (MeitY), Government of India, Grant: EE-9/2/21 - R\&D-E.}

}

\author{
    \IEEEauthorblockN{
    Tejas Chaudhari\IEEEauthorrefmark{1}\orcidlink{0009-0003-3317-1375},
    Akarsh J.\textdagger,\IEEEauthorrefmark{1}\orcidlink{0009-0000-0376-8304},
    Tanushree Dewangan\textdagger,\IEEEauthorrefmark{1}\orcidlink{0009-0009-3889-1228},\\
    Mukul Lokhande\IEEEauthorrefmark{1}\orcidlink{0009-0001-8903-5159}, 
    Santosh Kumar Vishvakarma\IEEEauthorrefmark{1}\orcidlink{0000-0003-4223-0077}, Senior Member, IEEE.}\\
    \IEEEauthorblockA{\IEEEauthorrefmark{1}NSDCS Research Group, Dept. of Electrical Engineering, Indian Institute of Technology Indore, India }\\
    Email: skvishvakarma@iiti.ac.in \textbf{(Corresponding Author)}
}

\maketitle

\begin{abstract}
This work proposes XR-NPE, a high-throughput Mixed-precision SIMD Neural Processing Engine, designed for extended reality (XR) perception workloads like visual inertial odometry (VIO), object classification, and eye gaze extraction. XR-NPE is first to support FP4, Posit (4,1), Posit (8,0), and Posit (16,1) formats, with layer adaptive hybrid-algorithmic implementation supporting ultra-low bit precision to significantly reduce memory bandwidth requirements, and accompanied by quantization-aware training for minimal accuracy loss. The proposed Reconfigurable Mantissa Multiplication and Exponent processing Circuitry (RMMEC) reduces dark silicon in the SIMD MAC compute engine, assisted by selective power gating to reduce energy consumption, providing 2.85$\times$ improved arithmetic intensity. XR-NPE achieves a maximum operating frequency of 1.72 GHz, area 0.016 mm\textsuperscript{2}, and arithmetic intensity 14 pJ at CMOS 28nm, reducing 42\% area, 38\% power compared to the best of state-of-the-art MAC approaches. The proposed XR-NPE-based AXI-enabled Matrix-multiplication co-processor consumes 1.4$\times$ fewer LUTs, 1.77$\times$ fewer FFs, and provides 1.2$\times$ better energy efficiency compared to SoTA accelerators on VCU129. The proposed co-processor provides 23\% better energy efficiency and 4\% better compute density for VIO workloads. XR-NPE establishes itself as a scalable, precision-adaptive compute engine for future resource-constrained XR devices. The complete set for codes for results reproducibility are released publicly, enabling designers and researchers to readily adopt and build upon them. \href{https://github.com/mukullokhande99/XR-NPE}{https://github.com/mukullokhande99/XR-NPE}.

\end{abstract}


\section{Introduction}
Recently, eXtended Reality (XR) technologies have
revolutionized user experience, with immersive domains such as Virtual Reality (VR), Audio Reality (AR) and Mixed-Reality (MR) offering captivating potential in exergaming, edutainment, etc\cite{XR, AR_VR_survey, AR-SoC, AI_accl_meta, Meta-VR_issue, MEGA.mini}. VR refers to simulated physical surroundings for users virtually, typically with projections with Head-mounted displays (HMD) and interaction control with joystick-like controllers or motion tracker handles. AR blends virtual content (information/interactive components) with real-sensory inputs (objects/gestures/voice) to enhance the user's perception of reality in smartphones, AI glasses or transparent headsets. MR merges both AR \& VR, creating human-computer inter-cohesion and/or real-time alignment of spatial objects in the physical world, with AI sensors, cameras or spatial mapping technologies.

\begin{figure}[!t]
    \centering
    \includegraphics[width=0.85\columnwidth]{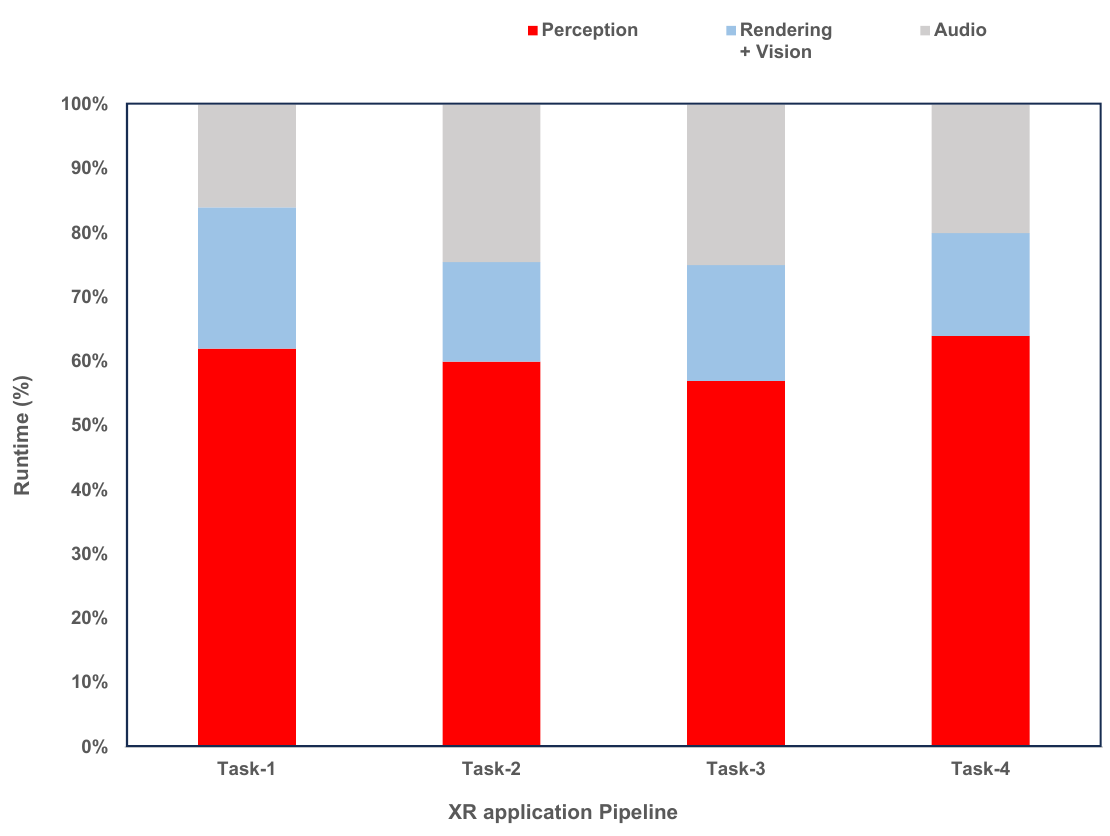}
    \caption{Workload analysis for application runtime \cite{Aspen}  emphasizing the computationally dominant perception workloads.}
    \label{fig:workload_char}
\end{figure}

\begin{table}[!b]
\caption{Qualitative Comparison of State-of-the-Art AI Accelerators and features in MAC Compute Engines.}
\label{tab:SOTA-comp-features}
\renewcommand{\arraystretch}{1.35}
\resizebox{\columnwidth}{!}{%
\begin{tabular}{|l|ll|ll|l|}
\hline
\multicolumn{1}{|c|}{\multirow{2}{*}{\textbf{Design}}} & \multicolumn{2}{c|}{\textbf{Precision}} & \multicolumn{2}{c|}{\textbf{Design}} & \multicolumn{1}{c|}{\multirow{2}{*}{\textbf{Use-cases}}} \\ \cline{2-5}
\multicolumn{1}{|c|}{} & \multicolumn{1}{c|}{\textbf{Datatype}} & \multicolumn{1}{c|}{\textbf{Bit-width}} & \multicolumn{1}{c|}{\textbf{Approach}} & \multicolumn{1}{c|}{\textbf{Overhead}} & \multicolumn{1}{c|}{} \\ \hline
\textbf{JSSC'25\cite{Occamy_JSSC}} & \multicolumn{1}{l|}{FP} & 8/16/32/64 & \multicolumn{1}{l|}{Radix-4 Booth} & \multicolumn{1}{c|}{-} & Server-class GPU \\ \hline
\textbf{TCAD'25\cite{AMD-MACC-TCAD'25}} & \multicolumn{1}{l|}{FP/BF16/TF32} & 4/8/16/32 & \multicolumn{1}{l|}{LUT} & \multicolumn{1}{c|}{Power} & Versal MPSoC \\ \hline
\textbf{TCAS-I'25\cite{Maestro}} & \multicolumn{1}{l|}{FP/BF16} & 8/16/32/64 & \multicolumn{1}{l|}{SPM-DOTP} & \multicolumn{1}{c|}{-} & AIoT \\ \hline
\textbf{TVLSI'25\cite{Flex-PE}} & \multicolumn{1}{l|}{Ad-FxP} & 4/8/16/32 & \multicolumn{1}{l|}{Unified CORDIC} & Energy Efficiency & Edge \\ \hline
\textbf{MICRO'24\cite{AMD-XDNA}} & \multicolumn{1}{l|}{INT/BF16} & 4/8/16/32 & \multicolumn{1}{c|}{-} & Memory Bound Compute & Mobile PCs - NPU \\ \hline
\textbf{HCS'24\cite{NVIDIA-blackwell}} & \multicolumn{1}{l|}{MXFP/BF16} & 4/6/8/16 & \multicolumn{1}{l|}{Mixed-precision} & \multicolumn{1}{c|}{-} & Blackwell GPU \\ \hline

\textbf{TCAS-II'24\cite{FMA-TCASII'24}} & \multicolumn{1}{l|}{FP/BF16/TF32} & 16/32/64 & \multicolumn{1}{l|}{BPD, CEP} & \begin{tabular}[c]{@{}l@{}}Area (Mant. mult)\\ Delay (Exp. proc.)\end{tabular} & \begin{tabular}[c]{@{}l@{}}High Performance\\ Computing\end{tabular} \\ \hline
\textbf{JSSC'23\cite{JSSC-Samsung}} & \multicolumn{1}{l|}{INT/FP} & 4/8/16 & \multicolumn{1}{l|}{\begin{tabular}[c]{@{}l@{}}Parallel Mult., \\ Adder Tree Acc.\end{tabular}} & \begin{tabular}[c]{@{}l@{}}Resources utilization\\ (Dark-Silicon)\end{tabular} & Smart-phone SoC \\ \hline
\textbf{ISCA'21\cite{RAPID-IBM}} & \multicolumn{1}{l|}{FP/FxP} & 2/4/8/16 & \multicolumn{1}{l|}{Approximation} & Accuracy Drop & - \\ \hline

\textbf{This Work} & \multicolumn{1}{l|}{FP/Posit} & 4/8/16 & \multicolumn{1}{l|}{SPM, PEP} & Run-time adjustable Perf. & XR Perception \\ \hline
\end{tabular}}
\end{table}

XR applications shall be executed on-device, especially with strict latency requirements and a need for continuous responsiveness, unreliability of network connectivity, and privacy concerns from cloud-execution. The detailed workload-analysis on application runtime discusses the major pipelines, such as the perception pipeline connected with sensors (camera/inertial measurement units) to extrapolate the information for enhanced understanding of user/physical environments (works at a speed greater than the sensor sampling rate: Camera - 30 fps), visual and audio pipelines to generate the information for the user. The workload characterization (Fig. \ref{fig:workload_char}) shows 60\% of  the total application-runtime is dominated by the perception pipeline, which consists primarily of computer vision/ DNN tasks. Aspen\cite{Aspen} highlighted the necessity to provide unified hardware support for the acceleration of different XR workloads, such as Eye-Gaze extraction to understand human sight/vision (DNNs), Visual-Inertial-Odometry (VIO) for user position (CV) and Object classification of scenario information (DNNs), etc.

\begin{figure*}[!t]
\centering
\begin{tabular}{ccc}
    \begin{subfigure}{}
        \centering
        \includegraphics[width=0.28\textwidth]{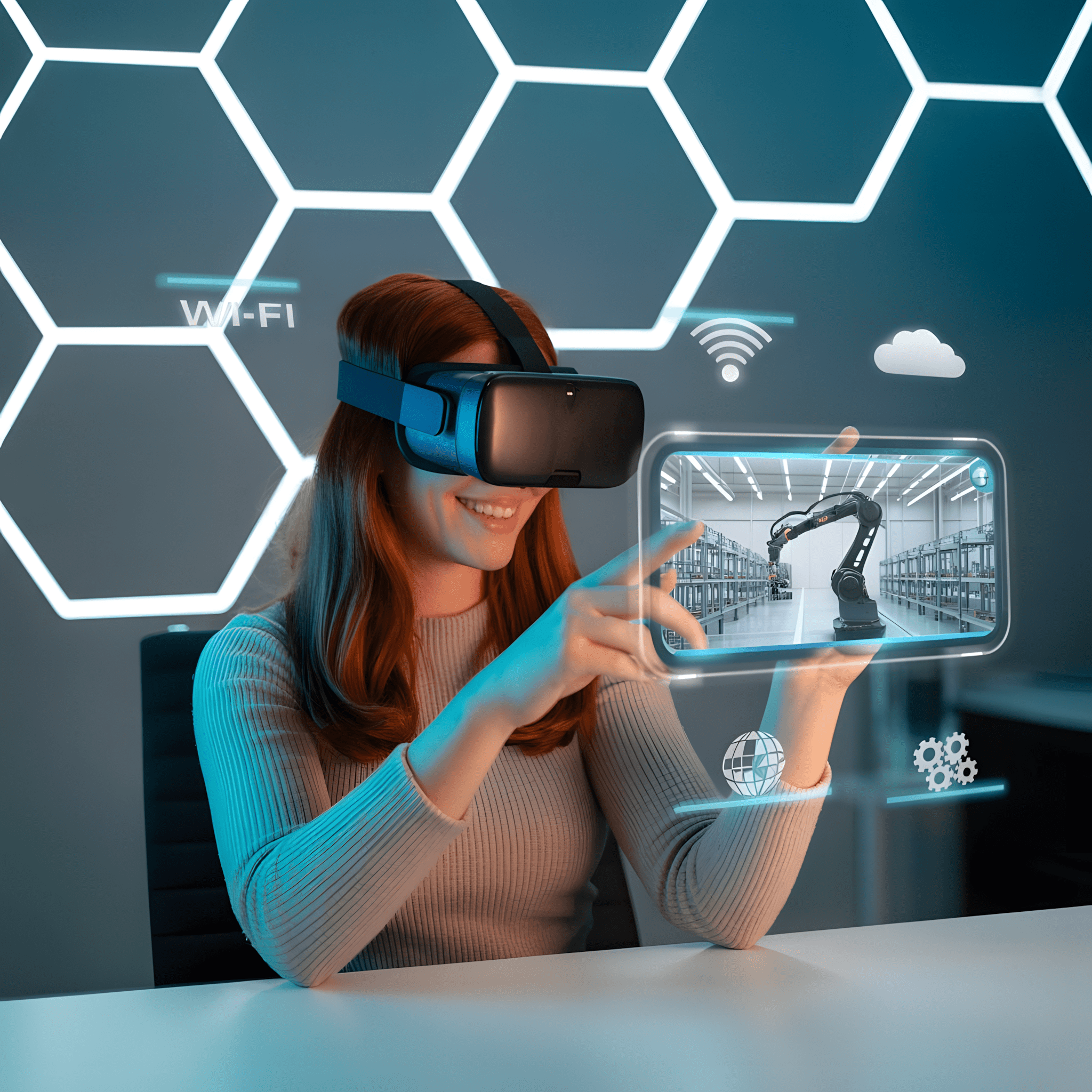}
        \label{fig:AR}
    \end{subfigure} &
    \hspace{2mm}
    \begin{subfigure}{}
        \centering
        \includegraphics[width=0.28\textwidth]{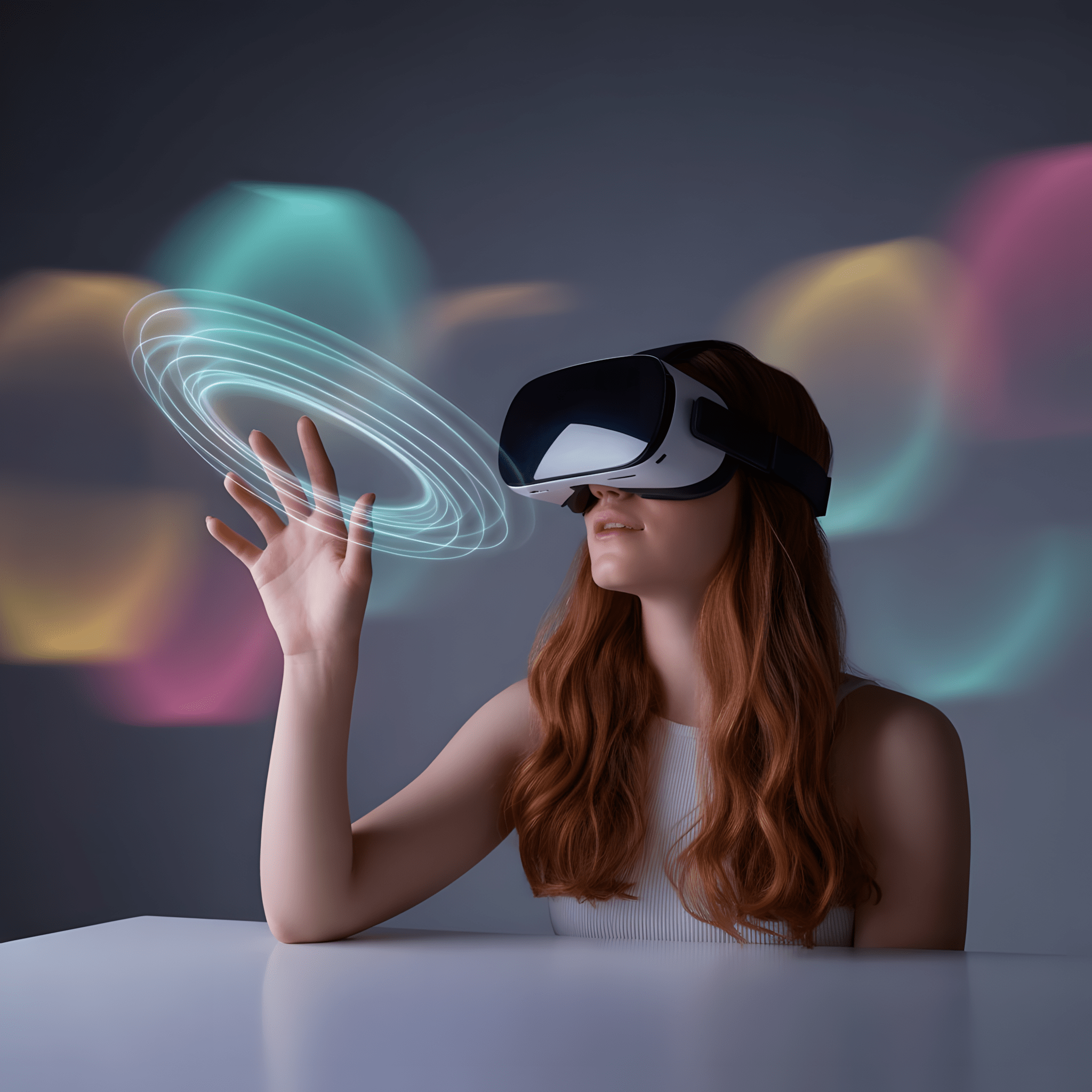}
        \label{fig:VR}
    \end{subfigure} &
    \hspace{2mm}
    \begin{subfigure}{}
        \centering
        \includegraphics[width=0.28\textwidth]{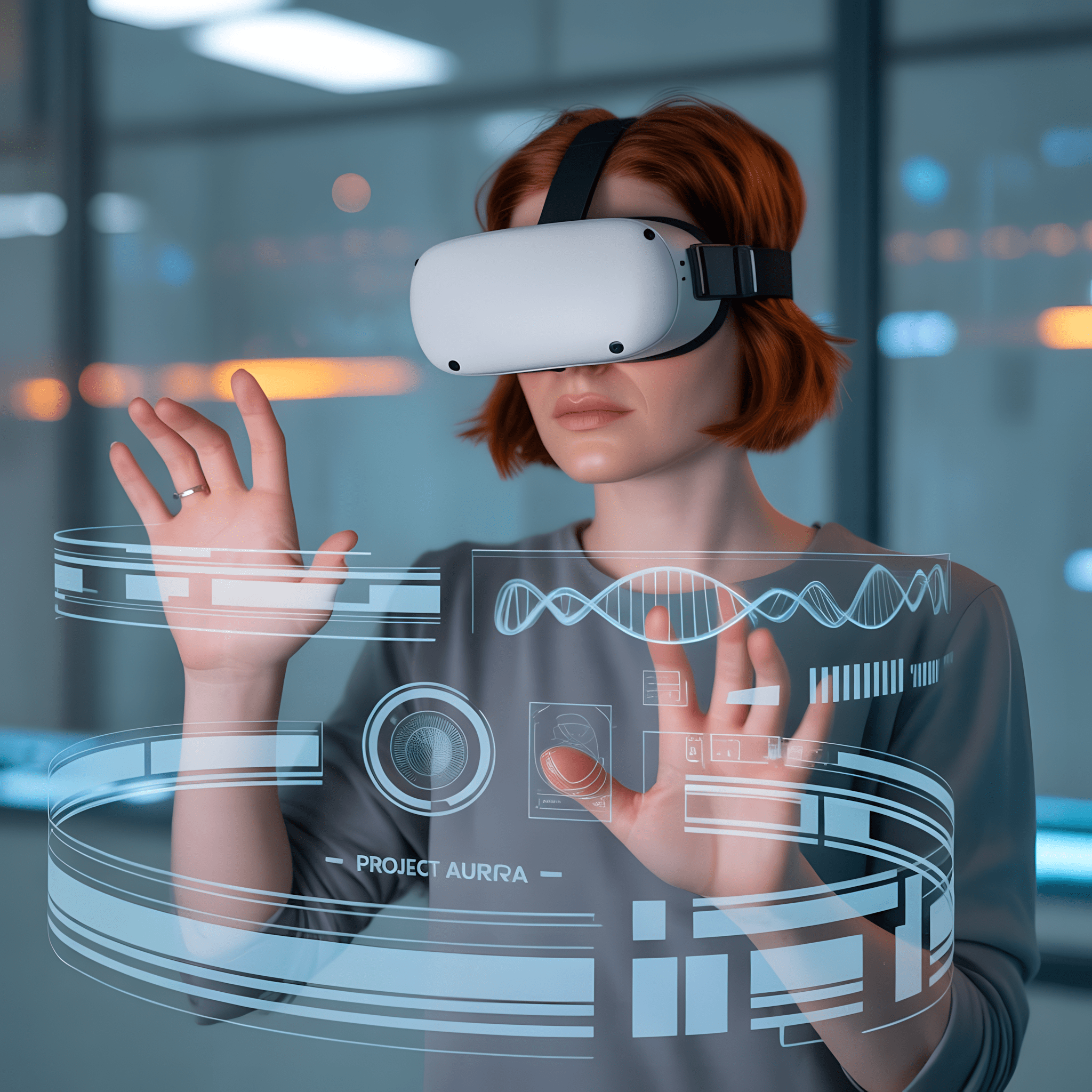}
        \label{fig:MR}
    \end{subfigure} \\
\end{tabular}
\caption{Comprehensive visualization for different eXtended Reality technologies, such as (a) Augmented-Reality, (b) Virtual-Reality, (c) Mixed-Reality. Image adapted from \cite{XR}.}
\end{figure*}

Prior XR SoC Designs for perceptions involved a dedicated VIO accelerator contributing up-to 72\% area in Meta's Oculus Quest 2 Mobile VR Headset\cite{AI_accl_meta}. Thus, an alternative that emerged would be the usage of existing DNN accelerators such as systolic array and vector engine for DNN-based VIO and significantly reduce dark-silicon. Furthermore, it shall also be noted\cite{Aspen} that the existing VIO models are too large in size for edge XR implementation and highly sensitive to quantization errors, leading to accuracy loss up-to 83\% (FP32 baseline). This work uses the UL-VIO model with the KITTI Odometry dataset, which has RGB images(1241×376). The model size reduces to 2.42 MB further with HFP4/Posit-4/Posit-8 mixed-precision approach, compared to 13.5MB (FP32), 3.4MB (FP8/INT8), 3.6MB\cite{Aspen} (Posit-8/16); albeit accuracy, model size tradeoff. The work further details the in-depth discussion on Mixed-precision quantization for different workloads such as DNNs, Object-classification, Eye-Gaze extraction and other models with Hybrid layer-adaptive quantized acceleration. The work also discusses XR-NPE hardware implementation and performance comparison with SoTA compute engines. Empirical analysis for the modular hybrid morphable matrix-array with XR-NPE is presented, along with SoTA comparison.

\begin{figure}[!b]
    \centering
    \includegraphics[width=0.85\columnwidth, height=80mm]{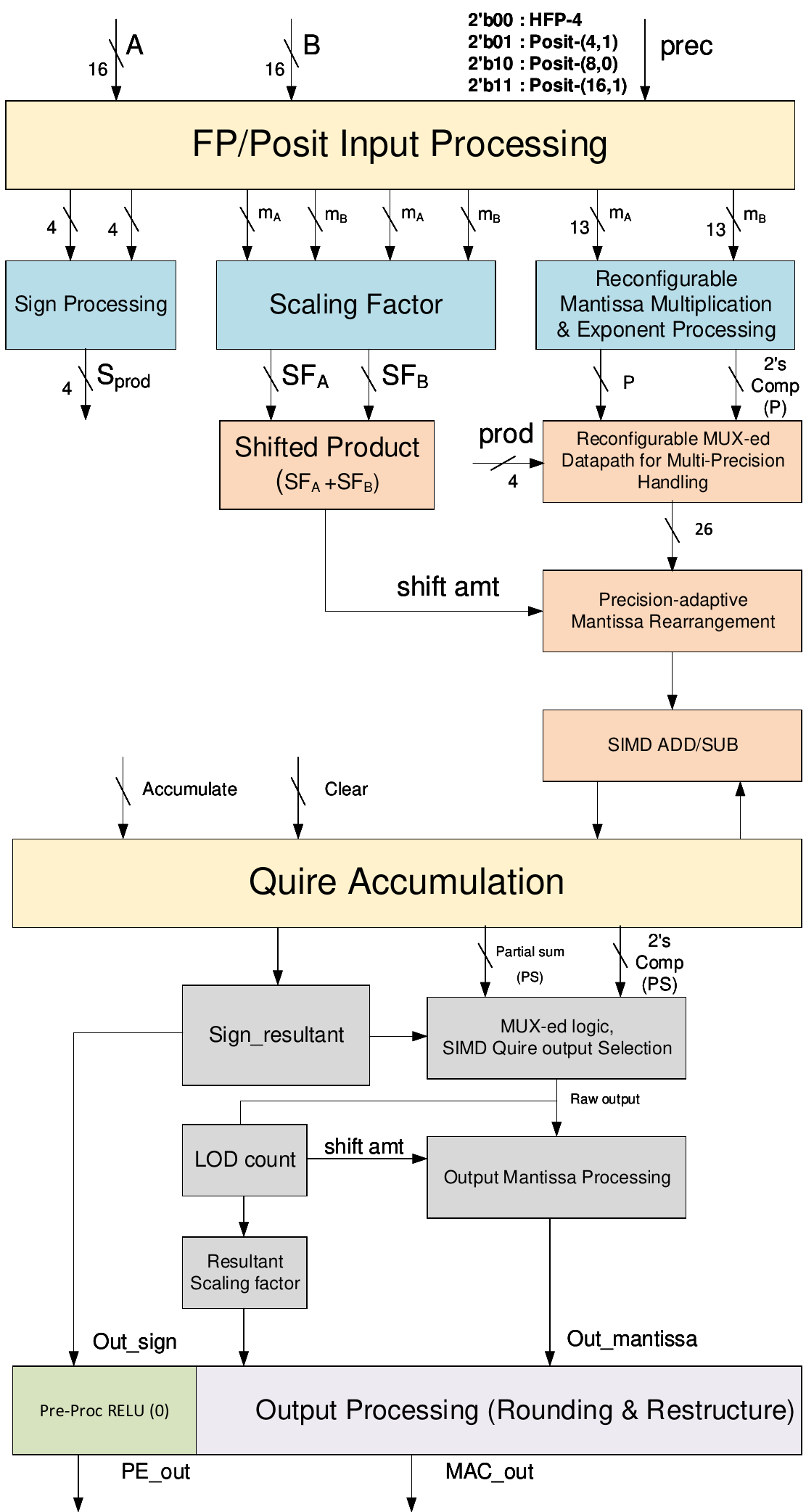}
    \caption{The proposed SIMD HFP-4, Posit(4,1), Posit(8,0), Posit(16,1) SIMD and Mixed-Precision XR-NPE, showcasing internal circuitry.}
    \label{fig:NPE}
\end{figure}

\section{Mixed-precision Neural Processing Engine}

Prior works have demonstrated SIMD NPE to be resource-efficient approach for mixed-precision computations\cite{LPRE, Flex-PE}. This has been found beneficial with hardware-algorithm co-design gains at edge implementations. The detailed microarchitecture for the proposed SIMD XR-NPE has been illustrated in the Fig. \ref{fig:NPE}. The modular approach follows different stages such as FP/Posit Input processing, Multiplication-stage, Quire scale-accumulate stage and Output processing. The proposed SIMD NPE supports 4x FP4/Posit(4,1) or 2x Posit(8,0) or 1x Posit(16,1) precision based on prec\_sel mode signal. The stage is also responsible for NaN, or normal-subnormal FP/posit, infinity,  and zero check on input operand and handles exception accordingly. 

The multiplication stage involves the calculation of the sign XOR, scaling factor for exponent/regime processing and a reconfigurable Mantissa-multiplication block. The unified comparator and Leading-one-detector are useful in a SIMD mixed-precision setup for smoother comparison. The key point to be noted is that adder/comparator hardware is linearly scalable, while shifter/multiplier hardware is exponentially scaled as we move from lower-precision to higher precision, thus marking significant dark-silicon in the multiplication stage, which can be reduced with RMMEC (reconfigurable mantissa multiplication, exponent processing circuitry) block. Our approach uses K-map based reconfigurable 2-bit RMMEC-block, Based on which we designed higher bit-width multiplication with low precision combination approach from 2-bit in Posit(4,1) / FP4 to 6-bit in Posit(8,0) and 12-bit in Posit(16,1). During zero input operands, the particular multiplier is power-gated and zero is fed to the accumulator accordingly. The approach uses a SIMD ADD/SUB block based on precision-adaptive rearrangement for quire accumulation. Followed by the output processing based on resultant sign, scaling factor and regime/exponent restructuring and mantissa rounding. 

\begin{figure}[!t]
    \centering
    \includegraphics[width=1\linewidth]{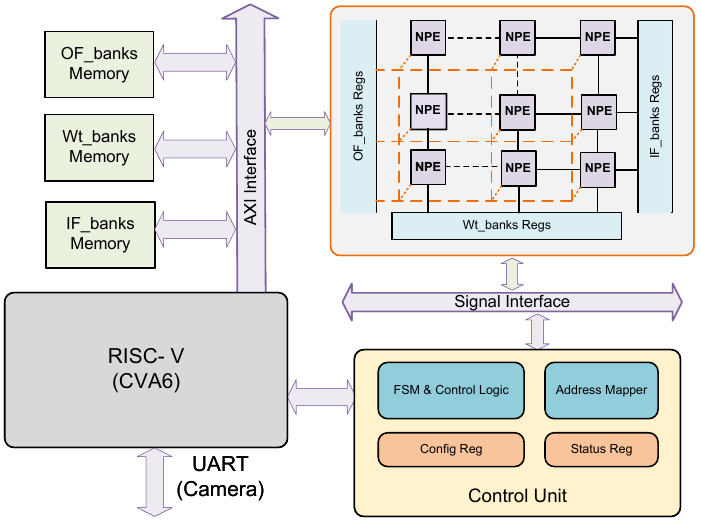}
    \caption{Detailed data flow with the proposed XR-NPE based
    Mixed-precision morphable matrix-multiplication array, RISC-V interface, memory banks, and control engine.}
    \label{fig:morph-array}
\end{figure}

The proposed evaluation system (Fig. \ref{fig:morph-array}) consists of an AXI-enabled mixed-precision morphable matrix-multiplication array, memory banks to feed input/output data, RISC-V interface, and control engine. The proposed co-processor accelerator is scalable (8x8 and 16x16) for evaluation. We have considered 8x8,  considering that other SoTA works assume 64 MAC units. The accelerator can be easily interfaced with AXI and DMA of Cheshire and assisted by validation via p-type SIMD ISA-based application programming interface (API)\cite{Flex-PE}. The proposed morphable matrix-multiplication accelerator plays  a key role in the execution of different AI workload, based on scheduling and control mechanisms as per workloads configurations for model layers, image size, layer type, kernel size, stride, type of pooling, and Activation function. Cheshire\cite{Cheshire} is a lightweight, linux-capable RISC-V host platform for the accelerator plug-in in the AMD Vivado Design Suite. The synchronization is managed via status/configuration registers, control signals, and a custom matrix-multiplication accelerator. The control units hold the details with Configuration/Status Registers, FSM Logic/Flags required for sequential computations and data-flow within the accelerator and the host processor. The proposed accelerator benefits from hardware–algorithm co-design techniques, for the above-mentioned applications such as object detection/classification, Eye-Gaze extraction, and Visual-Inertial Odometry models in a layer-adaptive mixed-precision approach for FP4/Posit(4,1)/Posit(8,0)/Posit(16,1). The proposed API was also demonstrated to have elevated performance for  most AI applications\cite{Posit-Transformers}.

\section{Performance Evaluation}

The experimental setup for the evaluation in the proposed work includes a quantized algorithmic analysis (emulated software framework) and hardware architecture design.

The layer-specific evaluations for the determination of the optimal precision for each layer are determined before inference itself. Quantization-aware training (QAT) is applied to ensure minimal error loss and is finalized. The proposed approach primarily analyzed the network with a particular layer in either of FP4/8/16/32, Posit-4/8/16/32, while keeping other layers in FP32, and evaluated the performance. The QAT model undergoes quantization while preserving accuracy degradation. The activations are retained with particular precision across all layers, while computations remain in FP-arithmetic, eliminating additional overhead during inference. Even after aggressive quantization, the retraining process maintains minimal accuracy loss. The overall loss function is based on the significance of this error across layers and their contributions. This is similar to~\cite{Quant4b-algo, HNPU-JSSC} , with a first-order Taylor expansion of the loss function and handles the layer parameters with the quantized model performance. This sensitivity metric enables selective low-bit quantization while maintaining minimal layers in higher precision. This dynamic adjustment approach also ensures model performance with efficient memory utilization and reduces computational complexity with power consumption.

\begin{figure}
    \centering
    \includegraphics[width=0.95\columnwidth]{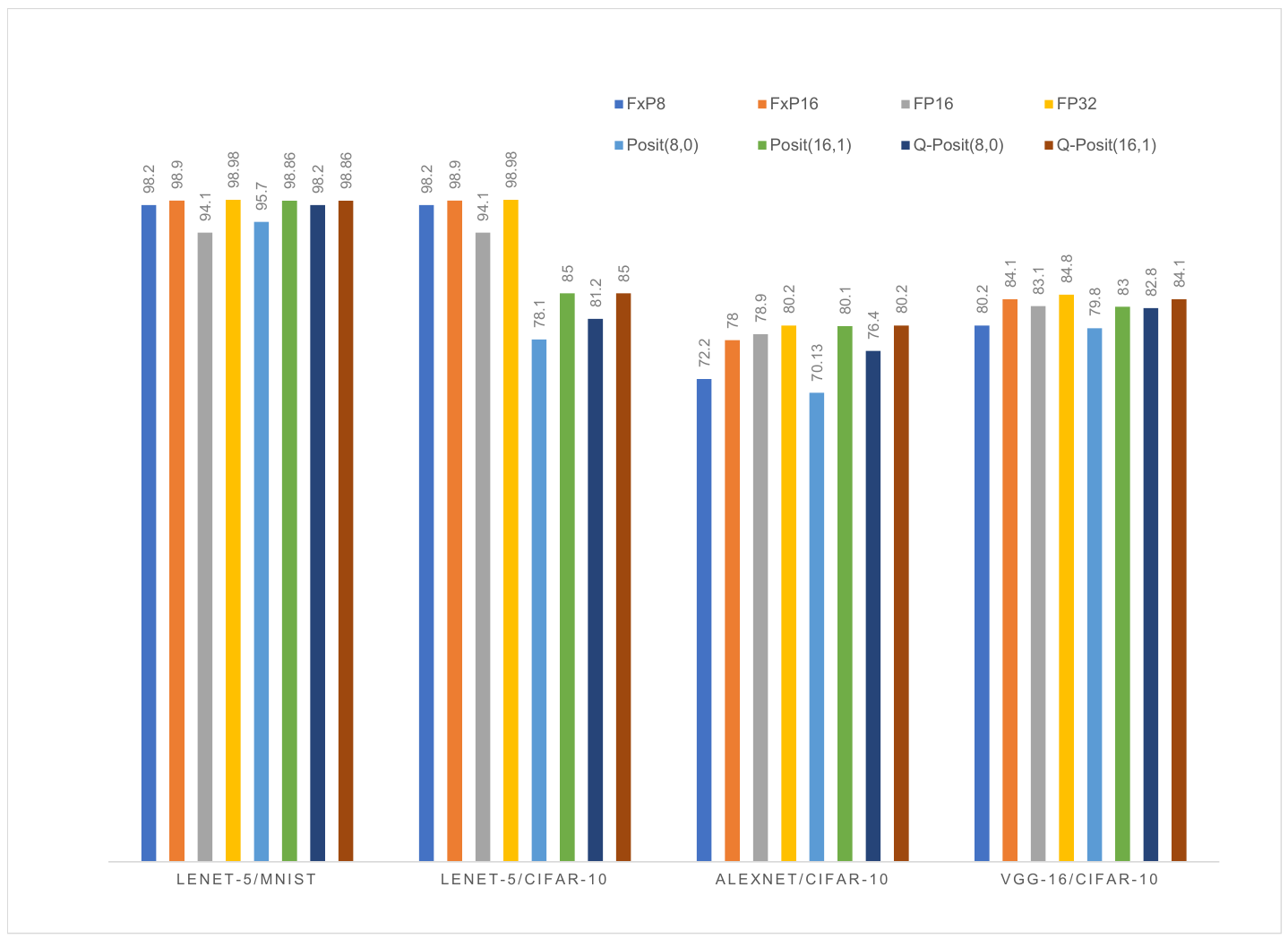}
    \caption{Comparative application accuracy for object classification model in different precision against FP32 baseline and SoTA FxP implementation\cite{Flex-PE}.}
    \label{fig:obj-accuracy}
\end{figure}

The sensitivity metric ($s$), denoted by $s_l$ the for $l$\textsuperscript{th} layer, is defined:
\begin{equation}
    s_{l_{\text{sc},k}} = \frac{\left( ||Q^{\text{MxP}}(\mathbf{w}_l) - \mathbf{w}_l|| - ||Q^{\text{MxP'}}_{\text{sc},k}(\mathbf{w}_l) - \mathbf{w}_l|| \right) \times ||\nabla \mathcal{L}_{\mathbf{w}_l}||}{n_l}
    \vspace{-2mm}
\end{equation}
with
\begin{equation}
    s_l = \max(s_{l_{\text{sc},8}}, s_{l_{\text{sc},4}}).
\end{equation}

\begin{figure}[!t]
    \centering
    \includegraphics[width=0.85\columnwidth]{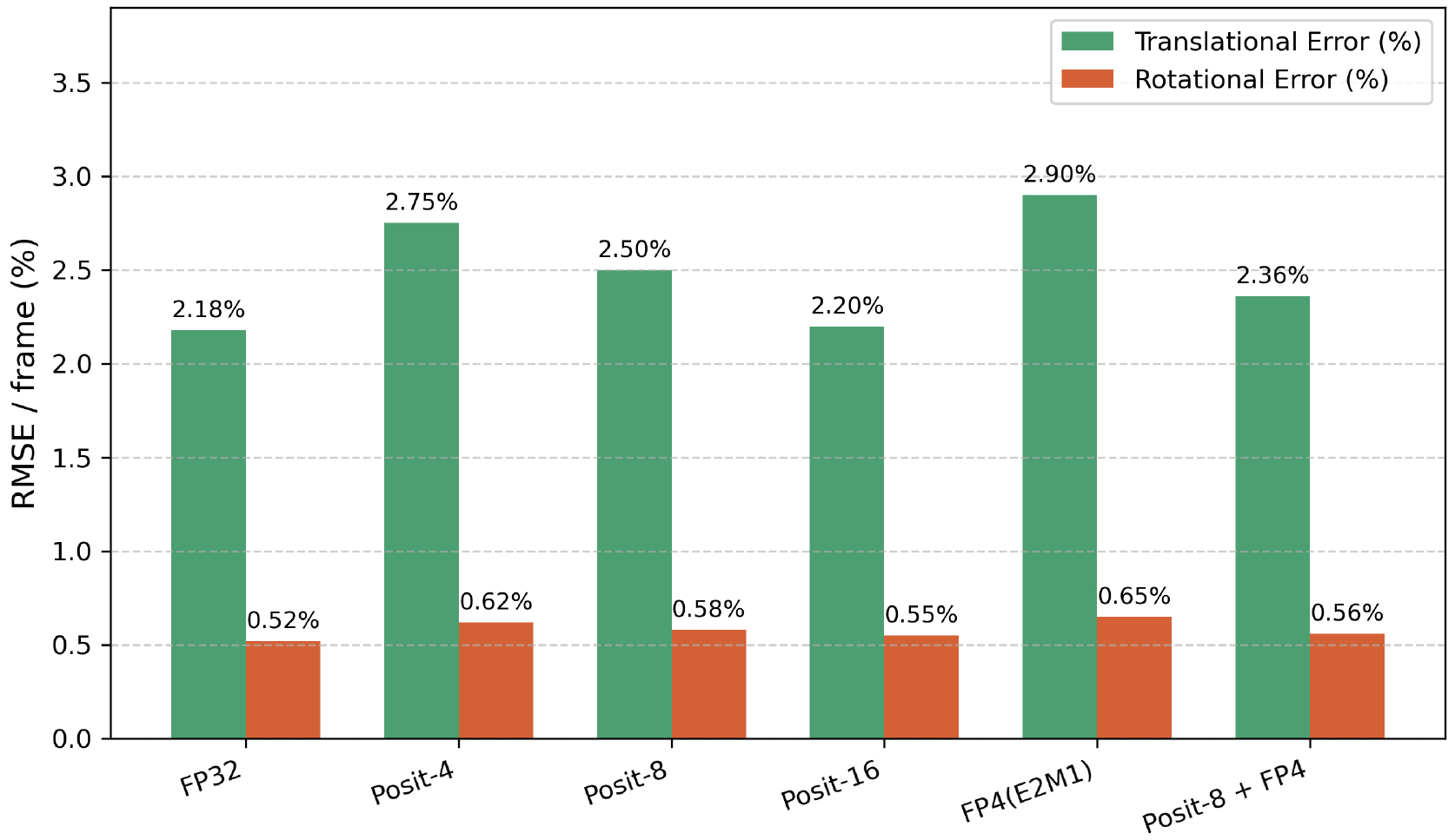}
    \caption{Precision-adaptive accuracy for Ultra Lightweight VIO model\cite{UL-VIO}.}
    \label{fig:UL-VIO}
\end{figure}

\begin{figure}[!t]
    \centering
    \includegraphics[width=0.95\columnwidth]{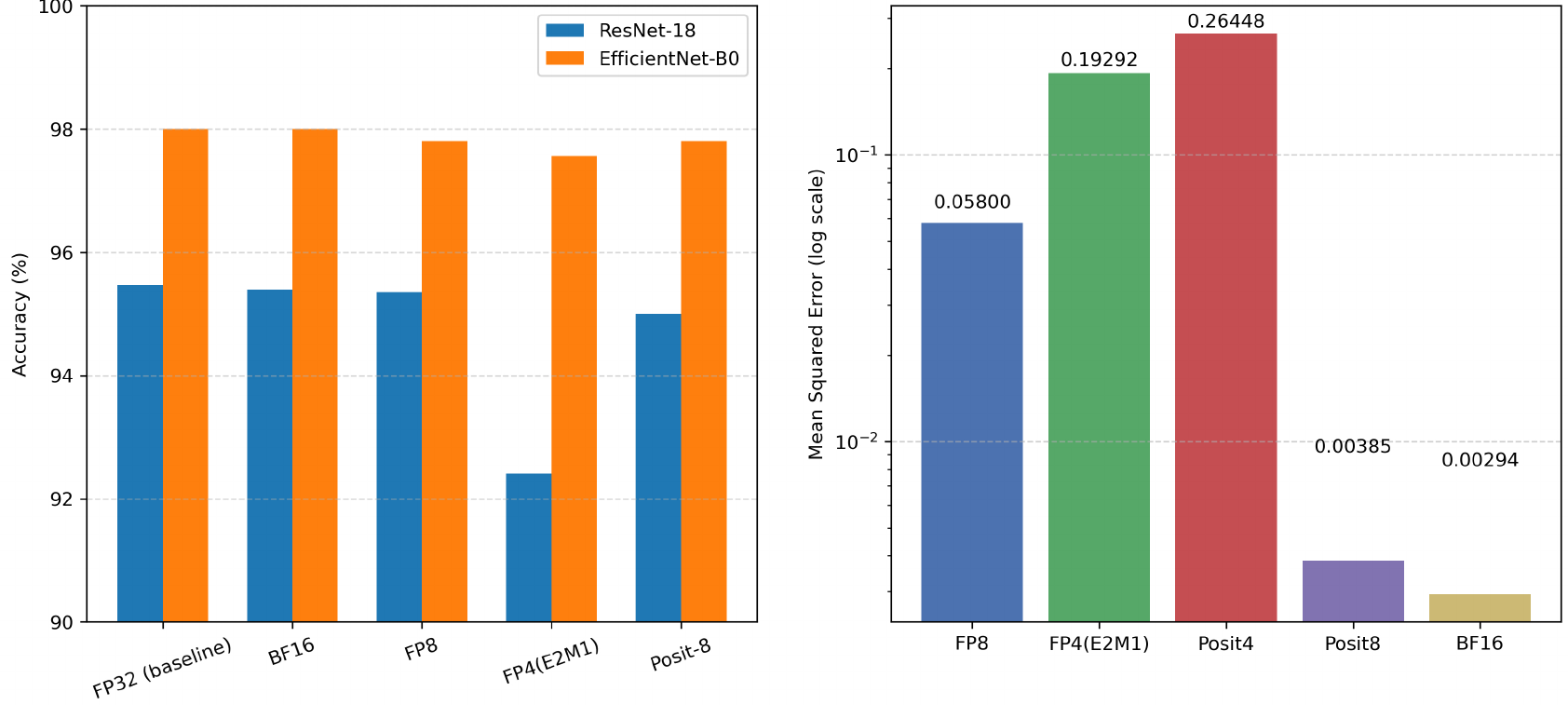}
    \caption{Impact of precision performance on application accuracy for object detection, eye-gaze LLE estimation.}
    \label{fig:MxP}
\end{figure}

The entropy-based uniform quantization scheme~\cite{Quant4b-algo} ensures an effective tradeoff between bit-width reduction and accuracy retention. The method uses a flexible combination of fractional and integral bit widths, dynamically adjusting lower[W\textsubscript{l}] and upper saturation thresholds [W\textsubscript{l}] to align with the model’s learned weight distribution, unlike conventional [-1,1], QMxP(.) as mixed-precision quantization, and W is the weight value.

\begin{equation}
    \text{scale (k)} = \text{mean}(\text{abs}(W)) \times \frac{2^n - 1}{2^{n-1}}
    \vspace{-3mm}
\end{equation}

\begin{equation}
    \widehat{W} = \text{round} \left( \left( \text{clip} \left( \frac{W}{\text{k}}, \text{W\textsubscript{l}}, \text{W\textsubscript{h}} \right) - \text{W\textsubscript{l}} \right) \times \frac{2^n - 1}{\text{W\textsubscript{h}} - \text{W\textsubscript{l}}} \right)
    \vspace{-1mm}
\end{equation}

\begin{equation}
    Q^{\text{MxP}}(W) = \widehat{W} * \frac{\text{W\textsubscript{h}} - \text{W\textsubscript{l}}}{2^n - 1} + \text{W\textsubscript{l}}
\end{equation}

The parameterized clipping activation (PACT) allows for accuracy loss recovery by training the clipped threshold. The uniform quantization around zero and the dynamically varied bit-width maintain the quantized values around zero and the balanced weight distribution. QAT is proven to compensate for approximation errors in reduced-bit precision operations. The layer-wise sensitivity~\cite{Quant4b-algo} prevents significant degradation in the critical layers. 

\begin{equation}
    \vspace{-2mm}
    y = \text{PACT}(x) = 0.5 (|x| - |x - \alpha| + \alpha).
    \vspace{-1mm}
\end{equation}

\begin{equation}
    x^q = \text{round} \left( y \times \frac{2^n - 1}{\alpha} \right) \times \frac{\alpha}{2^n - 1}.
\end{equation}

\begin{figure}[!t]
    \centering
    \includegraphics[width=0.95\columnwidth]{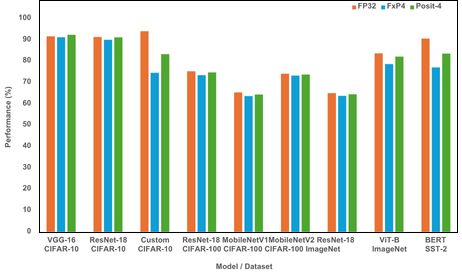}
    \caption{Comparative application accuracy for different AI models to be used in XR applications at different precision against FP32 baseline.}
    \label{fig:AI_m_XR}
\end{figure}

\begin{table}[!t]
\caption{ASIC Performance Comparison of the proposed Compute Engine with the SoTA SIMD MAC compute engines.}
\renewcommand{\arraystretch}{1.15}
\centering
\resizebox{\columnwidth}{!}{%
\label{tab:mac-asic}
\begin{tabular}{|l|c|c|c|c|c|c|}
\hline
\multirow{2}{*}{\textbf{Design}} & \textbf{Tech.} & \textbf{Voltage} & \textbf{Freq.} & \textbf{Area} & \textbf{Power} & \textbf{Arith. Intensity} \\ \cline{2-7} 
 & \textbf{nm} & \textbf{V} & \textbf{GHz} & \textbf{mm\textsuperscript{2}} & \textbf{mW} & \textbf{pJ/Op.} \\ \hline

\multirow{2}{*}{\textbf{TCAS-AI'25}\cite{FP_FMA_TCASAI25}} & \multirow{2}{*}{65} & \multirow{2}{*}{1.2} & 0.83 & 0.036 & 29.68 & 142.5 \\\cline{4-7}
 &  &  & 0.74 & 0.0395 & 33.80 & 183 \\\hline
\textbf{TCAS-I'25}\cite{FMA-TCASI'25} & 28 & 1 & 0.97 & 0.0276 & 39 & 40 \\\hline
\textbf{TVLSI'25}\cite{Flex-PE} & 28 & 0.9 & 1.36 & 0.049 & 7.3 & 5.37 \\\hline
\textbf{TCAS-II'24}\cite{FMA-TCASII'24} & 28 & 1 & 1.56 & 0.022 & 72.3 & 46.35\\ \hline
\textbf{TCAD'24}\cite{DPDAC-TCAD'24} & 28 & 1 & 1.47 & 0.024 & 82.4 & 56\\ \hline
\textbf{TCAS-II'22}\cite{UVMAC-TCASII'22} & 28 & 1.05 & 0.67 & 0.052 & 99 & 148\\ \hline
\textbf{This work} & 28 & 0.9 & 1.72 & 0.016 & 24.1 & 14 \\ \hline
\end{tabular}}
\end{table}

\begin{table}[!t]
\caption{Comparative analysis with diverse SoTA accelerator approaches}
\label{tab:fpga-arch-comp}
\renewcommand{\arraystretch}{1.25}
\resizebox{\columnwidth}{!}{%
\begin{tabular}{|l|l|c|c|c|c|c|}
\hline
\textbf{Parameters} & \multicolumn{1}{c|}{\textbf{This work}} & \textbf{TVLSI'25\cite{Flex-PE}} & \textbf{TCAS-II'23\cite{Ski-TCASII'23}} & \textbf{ISCAS'25\cite{LPRE}} & \textbf{TCAS-I'24\cite{MKim-TCAS-I'24}} & \textbf{TCAS-I'24\cite{BWu-TCAS-I'24}} \\ \hline
\begin{tabular}[c]{@{}l@{}}\textbf{MPSoC}\\\textbf{Evaluation Kit}\end{tabular} & \begin{tabular}[c]{@{}c@{}}XCZU7EV-\\ 2FFVC1156\end{tabular} & \begin{tabular}[c]{@{}c@{}}XCVU29P-\\ L2FSGA2577E\end{tabular} & \begin{tabular}[c]{@{}c@{}}XCVU9P-\\ 2FLGA2577I\end{tabular} & \begin{tabular}[c]{@{}c@{}}XC7Z020-\\ 1CLG400C\end{tabular} & XC7A100T & \begin{tabular}[c]{@{}l@{}}XAZU3EG-\\ 1SFVC784I\end{tabular} \\ \hline
\textbf{Tech. (nm)} & \multicolumn{1}{c|}{16} & 16 & 14 & 28 & 28 & 16 \\ \hline
\textbf{Model} & \multicolumn{1}{c|}{VIO} & VGG-16 & YOLO v3-Tiny & YOLO v3-Tiny & YOLO v3-Tiny & ResNet-50 \\ \hline
\textbf{Freq. (MHz)} & \multicolumn{1}{c|}{250} & 466 & 150 & 50 & 100 & 150 \\ \hline
\textbf{Bit-width} & \multicolumn{1}{c|}{4/8/16} & 4/8/16/32 & 8 & 8/16 & 8 & 8 \\ \hline
\textbf{LUTs (K)} & \multicolumn{1}{c|}{28.94} & 36.5 & 132 & 17.54 & 50.2 & 40.78 \\ \hline
\textbf{FFs (K)} & \multicolumn{1}{c|}{25.6} & 7.3 & 39.5 & 14.8 & 58.1 & 45.25 \\ \hline
\textbf{DSP blocks} & \multicolumn{1}{c|}{0} & 62 & 96 & 39 & 240 & 257 \\ \hline
\textbf{Power (W)} & \multicolumn{1}{c|}{1.2} & 1.72 & 5.52 & 0.93 & 2.2 & 1.4 \\ \hline
\textbf{\begin{tabular}[c]{@{}l@{}}Energy Efficiency\\ (GOPS/W)\end{tabular}} & \multicolumn{1}{c|}{53.4} & 10.96 & 6.36 & 2.14 & 43 & 45 \\ \hline
\end{tabular}}
\end{table}

We have compared precision-wise performance  (Fig. \ref{fig:UL-VIO}, Fig. \ref{fig:MxP}) for object classification (Efficient-Net), gaze estimation, and VIO model (KITTI dataset\cite{KITTI}). FP4 achieves near-BF16/FP8 classification accuracy and acceptable gaze MSE while significantly reduced bit-widh, which leads to significantly reduced hardware resources and power consumption associated with off-chip data-movement. It was found that off-chip data-movement accounts to almost 60\% of energy-consumption and latency in system-level architectural performance. The UL-VIO detailed, FP4 enhances translation and rotation RMSE by just 0.72 pp and 0.13 pp, respectively, compared to FP32 (baseline), while Posit-8/Posit-16 maintains high accuracy and minimal VIO errors. Ours proposed approach with MxP (Posit-8/FP4) achieves trade-off  combination compared to FP4 errors and Posit-8 hardware performance. The proposed findings shows significant enhancement with 4-bit for extreme model compression and Posit's precision range robustness, with mixed-precision scheme as optimal solution in resource-constrained environment.

During hardware evaluation, we synthesized the proposed XR-NPE and the AXI-enabled matrix-multiplication co-processor with CMOS 28-nm technology, followed by the place-and-route parameters being reported. We found the proposed design stands out in terms of operating frequency compared to SoTA MAC Compute designs, with significantly reduced arith. Intensity up to 2.85 $\times$, which reflects enhanced energy efficiency. This has been detailed in Table \ref{tab:mac-asic}, where XR-NPE is also found reasonably resource-efficient in area and power consumption, approximately 42\% reduced area and 38\% reduced power compared to the recent design\cite{FMA-TCASI'25}. We composed our co-processor architecture with FPGA design in AMD Vivado Design Suite, and the hardware performance comparison with SoTA accelerator approaches has been detailed in Table \ref{tab:fpga-arch-comp}. We compared iso-computational units (64) accelerator design and found the proposed approach consumes, 1.4$\times$ less LUTs and 1.77$\times$ less FFs and improved energy efficiency upto 1.2$\times$, compared to \cite{BWu-TCAS-I'24} on VCU129. 

\begin{table*}[!t]
\caption{Performance Metrics Comparison with SoTA AI co-processors.}
\centering
\renewcommand{\arraystretch}{1.25}
\resizebox{\textwidth}{!}{%
\label{tab:asic-arch-comp}
\begin{tabular}{|c|ccc|cccccc|}
\hline
\multirow{2}{*}{\textbf{Design}} & \multicolumn{3}{c|}{\textbf{Architectural Emulation}} & \multicolumn{6}{c|}{\textbf{Hardware Performance}} \\ \cline{2-10} 
 & \multicolumn{1}{c|}{\textbf{Network Topology}} & \multicolumn{1}{c|}{\textbf{Precision}} & \textbf{Accuracy (\%)} & \multicolumn{1}{c|}{\textbf{Tech. (nm)}} & \multicolumn{1}{c|}{\textbf{Freq. (MHz)}} & \multicolumn{1}{c|}{\textbf{Power (W)}} & \multicolumn{1}{c|}{\textbf{Area (mm\textsuperscript{2})}} & \multicolumn{1}{c|}{\textbf{\begin{tabular}[c]{@{}c@{}}Energy Efficiency\\ (TOPS/W)\end{tabular}}} & \textbf{\begin{tabular}[c]{@{}c@{}}Compute Density\\ (TOPS/mm\textsuperscript{2})\end{tabular}} \\ \hline
\multirow{2}{*}{\textbf{JSSC'25\cite{VSA-JSSC'25}}} & \multicolumn{1}{c|}{Vector Systolic Array} & \multicolumn{1}{c|}{\multirow{2}{*}{FxP4/8}} & 71.68 & \multicolumn{1}{c|}{\multirow{2}{*}{28}} & \multicolumn{1}{c|}{172} & \multicolumn{1}{c|}{0.6} & \multicolumn{1}{c|}{1.04} & \multicolumn{1}{c|}{8.33} & 7.94 \\ \cline{2-2} \cline{4-4} \cline{6-10} 
 & \multicolumn{1}{c|}{G-VSA} & \multicolumn{1}{c|}{} & 67.2 & \multicolumn{1}{c|}{} & \multicolumn{1}{c|}{199} & \multicolumn{1}{c|}{0.3} & \multicolumn{1}{c|}{2} & \multicolumn{1}{c|}{3.26} & 1.13 \\ \hline
\multirow{2}{*}{\textbf{TVLSI'25\cite{MSDF-MAC}}} & \multicolumn{1}{c|}{784-200-100-10} & \multicolumn{1}{c|}{\multirow{2}{*}{FxP8}} & 97.4 & \multicolumn{1}{c|}{\multirow{2}{*}{45}} & \multicolumn{1}{c|}{\multirow{2}{*}{588}} & \multicolumn{1}{c|}{0.61} & \multicolumn{1}{c|}{6.13} & \multicolumn{1}{c|}{1.48} & 0.144 \\ \cline{2-2} \cline{4-4} \cline{7-10} 
 & \multicolumn{1}{c|}{784-256-10} & \multicolumn{1}{c|}{} & 96.73 & \multicolumn{1}{c|}{} & \multicolumn{1}{c|}{} & \multicolumn{1}{c|}{0.64} & \multicolumn{1}{c|}{5.88} & \multicolumn{1}{c|}{1.39} & 0.153 \\ \hline
\textbf{JSSC'24\cite{Marsellus_JSSC}} & \multicolumn{1}{c|}{ResNet-20} & \multicolumn{1}{c|}{FP-16/32, BF16} & 92.2 & \multicolumn{1}{c|}{22} & \multicolumn{1}{c|}{420} & \multicolumn{1}{c|}{0.123} & \multicolumn{1}{c|}{1.9} & \multicolumn{1}{c|}{12.4} & - \\ \hline
\textbf{TCAS-I'22\cite{PL-NPU_TCASI'22}} & \multicolumn{1}{c|}{ResNet-18} & \multicolumn{1}{c|}{Posit-8} & 70.1 & \multicolumn{1}{c|}{28} & \multicolumn{1}{c|}{1040} & \multicolumn{1}{c|}{343} & \multicolumn{1}{c|}{5.28} & \multicolumn{1}{c|}{1.63} & 0.101 \\ \hline
\textbf{ISCAS'24\cite{ODL-ISCAS'24}} & \multicolumn{1}{c|}{ResNet-50} & \multicolumn{1}{c|}{FxP4/FP-16/32} & 77.56 & \multicolumn{1}{c|}{28} & \multicolumn{1}{c|}{160} & \multicolumn{1}{c|}{67.4} & \multicolumn{1}{c|}{1.84} & \multicolumn{1}{c|}{2.19} & 0.085 \\ \hline
\textbf{This work} & \multicolumn{1}{c|}{EfficientNet} & \multicolumn{1}{c|}{FP-4 / Posit-4/8/16} & \multicolumn{1}{c|}{97.56} & \multicolumn{1}{c|}{28} & \multicolumn{1}{c|}{250} & \multicolumn{1}{c|}{4.2} & \multicolumn{1}{c|}{1.95} & \multicolumn{1}{c|}{15.23} & \multicolumn{1}{c|}{8.2} \\ \hline
\end{tabular}}
\end{table*}

The proposed co-processor has been compared with state-of-the approaches, and our work which follows DNN-based VIO was found to be superior in terms of both energy efficiency by 23 \% and compute density by 4\% compared to the best of prior work. Our work also provided a mixed-precision approach with SoTA low-bit precision, such as FP4 and Posit(4,1), which has been found to reduce memory-bandwidth and associated energy cost over existing FP8/BF16 implementation. We mark the further exploration into this as potential future work. 

\section{Conclusion \& Future Work}

This work presents XR NPE, the first runtime reconfigurable Mixed-precision SIMD neural processing engine with low-bit precision support for FP4/Posit-(4,1) in a layer-adaptive quantization fashion for XR-perception workloads. The proposed XR-NPE benefits from an RMMEC-based approach with a significant reduction of 42\% area, 38\% power compared to the best of state-of-the-art MAC approaches, leading to 2.85$\times$ improved arithmetic intensity. XR-NPE-based Matrix-Multiplication coprocessor confirms reduced resource consumption up to 1.4$\times$ LUT and 1.77$\times$ FF. The proposed coprocessor also surpasses prior ASIC accelerators with an improvement of 23\% energy efficiency and 4\% compute density. The proposed advancement enables real-time energy-efficient VIO workloads without significantly degrading application performance. This makes XR-NPE  a viable option for future resource-constrained XR platforms such as augmented and virtual reality headsets.

\bibliographystyle{ieeetr}
\bibliography{bib}

\end{document}